\documentclass[apj]{emulateapj}

\usepackage{natbib}
\usepackage{multirow}
\usepackage{appendix}
\usepackage{longtable}

\slugcomment{ApJ Accepted}

\shorttitle{NIR Polarimetry of ``Polar Scattered'' Seyfert 1's}
\shortauthors{Batcheldor et al.}

\begin{document}


\title{\sc{Nicmos} Polarimetry of ``Polar Scattered'' Seyfert 1 Galaxies\altaffilmark{1}}


\author{D. Batcheldor,\altaffilmark{2,4} A. Robinson,\altaffilmark{3,5} D. J. Axon,\altaffilmark{3,5,8} 
S. Young,\altaffilmark{3,5}, S. Quinn,\altaffilmark{3} J. E. Smith,\altaffilmark{6}, J. Hough,\altaffilmark{5} \& D. M. Alexander\altaffilmark{7}}
\email{dbatcheldor@fit.edu}




\altaffiltext{1}{Based on observations made with the NASA/ESA Hubble Space Telescope obtained at the 
Space Telescope Science Institute, which is operated by the Association of Universities for Research in Astronomy, Incorporated, under 
NASA contract NAS 5-26555. These observations are associated with program \#10160.}

\altaffiltext{2}{Department of Physics and Space Sciences, Florida Institute of Technology, 150 W. University Blvd, Melbourne, FL, 32901, USA}
\altaffiltext{3}{Department of Physics, Rochester Institute of Technology, 54 Lomb Memorial Drive, Rochester, NY, 14623, USA}
\altaffiltext{4}{Center for Imaging Science, Rochester Institute of Technology, 54 Lomb Memorial Drive, Rochester, NY, 14623, USA}
\altaffiltext{5}{Centre for Astrophysics Research, Science \& Technology Research Institute, University of Hertfordshire, Hatfield AL10 9AB, UK}
\altaffiltext{6}{The Open University, Walton Hall, Milton Keynes, MK7 6AA, UK}
\altaffiltext{7}{Department of Physics, Durham University, Durham DH1 3LE, UK}
\altaffiltext{8}{School of Mathematical \& Physical Sciences, University of Sussex, Falmer, Brighton, BN2 9BH, UK}



\begin{abstract}
The nuclei of Seyfert 1 galaxies exhibit a range of optical polarization characteristics that can be understood in terms of two scattering
regions producing orthogonal polarizations: an extended polar scattering region (PSR) and a compact equatorial scattering region 
(ESR), located within the circum-nuclear torus. Here we present NICMOS 2.0\micron\ imaging polarimetry of 6 ``polar scattered'' 
Seyfert~1 (S1) galaxies, in which the PSR dominates the optical polarization. The unresolved nucleus  ($<0\farcs58$) is significantly 
polarized in only three objects, but 5 of the 6 exhibit polarization in a 0\farcs58--1\farcs5 circum-nuclear annulus. In Fairall\,51 and 
ESO\,323-G077, the polarization position angle at 2\micron\ ($\theta_{2\micron}$) is consistent with the average for the optical spectrum
($\theta_v$), implying that the nuclear polarization is dominated by polar scattering at both wavelengths. The same is probably true for
NGC\,3227. In both NGC\,4593 and Mrk\,766, there is a large difference between $\theta_{2\micron}$ and $\theta_v$ off nucleus, where 
polar scattering is expected to dominate. This may be due to contamination by interstellar polarization in NGC\,4593, but there is no clear 
explanation in the case of the strongly polarized Mrk\,766. Lastly, in Mrk\,1239,  a large change ($\approx 60\degr$)
in  $\theta_{2\micron}$ between the nucleus and the annulus indicates that the unresolved nucleus and its immediate surroundings have
different polarization states at 2\micron, which we attribute to the ESR and PSR, respectively. A further implication is that the source of 
the {\em scattered} $2\micron$ emission in the unresolved nucleus is the accretion disk, rather than torus hot dust emission.
\end{abstract}




\section{Introduction}

In the Unified Model for Seyfert galaxies \citep{1993ARA&A..31..473A,1995PASP..107..803U}, Seyfert Type 1 (S1) and Type 2 (S2) nuclei are intrinsically the 
same type of object viewed at different orientations. In the S2's, our direct line-of-sight to the nuclear continuum source and broad-line region (BLR) is blocked by 
a clumpy toroidal region of dusty molecular gas clouds, on scales of possibly just a few parsecs \citep{2004Natur.429...47J, 2007A&A...474..837T}.
 Spectropolarimetry played a pivotal role in 
establishing this picture through the detection of polarized broad-lines. These features, attributed to scattering of broad-line emission above the poles of the torus 
(e.g., \citealt{1985ApJ...297..621A}), reveal the presence of an otherwise obscured BLR in many S2's. The search for polarized broad-lines has motivated many             
subsequent spectropolarimetric studies of S2's \citep{1992ApJ...397..452T,1996MNRAS.281.1206Y,2001ApJ...554L..19T}. 
The optical polarization position angle ($\theta_v$) in S2's is usually oriented perpendicular to the projected radio source axis and hence the axis of the obscuring 
toroidal region \citep{1983Natur.303..158A,1990MNRAS.244..577B}. This is consistent with the simple polar scattering envisaged in the unified model, because scattered 
light is polarized perpendicular to the scattering plane that contains the incident rays. 

\begin{deluxetable*}{lcccccccc}
\tablecaption{Sample of Polar-scattered Seyfert 1 Galaxies \label{tab:sample}}
\tablewidth{0pt}
\tablehead{
\colhead{Target}&
\colhead{Type}&
\colhead{$v_r$}&
\colhead{$\phi$}&
\colhead{H$\alpha$/H$\beta$}&
\colhead{$p_v$}&
\colhead{$\theta_{v}$}&
\colhead{RPA}&
\colhead{Refs.}\\
\colhead{}&
\colhead{}&
\colhead{(km~s$^{-1}$)}&
\colhead{(pc)}&
\colhead{}&
\colhead{(\%)}&
\colhead{(\degr)}&
\colhead{(\degr)} &
\colhead{}
}
\startdata
Fairall 51          & SB(rs)b Sy1                     & 4114 & 55 & $5.1\pm0.6$ & 3.9 -- 7.0 & 140 & \nodata & 1   \\
NGC 4593        & SB(rs)b Sy1                     & 2830 & 38 & $3.5\pm0.2$ & 0.1 -- 0.6 & 145  & \nodata & 1   \\
NGC 3227        & SAB(s) pec Sy1.5            & 1352 & 18 & $4.3\pm0.5$ & 0.6 -- 1.6 & 125  & 170     & 1,2 \\
Mrk 766            & SB(s)a: Sy1.5                   & 4104 & 55 & $6.1\pm0.6$ & 2.3 -- 3.9 & 90 & 27      & 3,4 \\
Mrk 1239          & E-S0  Sy1.5                       & 5941 & 79 & $4.7\pm0.8$ & 4.3 -- 6.4 & 129   & \nodata & 3   \\
ESO 323-G077 & SB(1)0 $\hat{}$ 0 Sy1.2 & 4417 & 59 & $5.9\pm1.7$ & 2.7 -- 4.7 & 84  & \nodata & 5   \\
\enddata
\tablecomments{
Properties of the optically selected sample of polar-scattered S1's. Types have been taken from NED\footnote{http://nedwww.ipac.caltech.edu/}. 
Virgo-infall corrected recessional velocities ($v_r$) have also been taken from NED and used to calculate the spatial scale of the {\it HST}+NICMOS
diffraction limit ($\phi$) at 2.0\micron\ (0\farcs2) assuming a Hubble constant of 73km~s$^{-1}$~Mpc. We have estimated H$\alpha$/H$\beta$ flux ratios 
from our optical spectropolarimetry of the broad lines (and in the case of ESO 323-G077, from the spectrum published by \citealt{2003A&A...404..505S}). 
The optical polarimetry covers the wavelength range 4500 -- 7100\AA; the range in percentage polarization (red-to-blue) and the average polarization position angle 
over this range are listed. The uncertainties on $\theta_V$ are all $\le1\degr$. RPA is the radio position angle. The 
most up-to-date references for the optical polarimetry (listed first) and radio data (listed second) are [1] \cite{2004MNRAS.350..140S}, 
[2] \cite{1995MNRAS.275...67M}, [3] Robinson et al. 2010 (in prep) [4] \cite{1999ApJS..120..209N} \& [5] \cite{2003A&A...404..505S}.
}
\end{deluxetable*}

In contrast, $\theta_v$ is parallel to the radio source axis in the majority 
of S1's \citep{1983Natur.303..158A,2002MNRAS.335..773S}. This implies scattering in a plane perpendicular to the system principal axis, 
which in turn indicates a second scattering region that is present in S1's but not observed in S2's. Spectropolarimetric studies have shown that S1's often exhibit 
distinctive structure in both $\theta_v$ and the percentage of optical polarization ($p_v$) across the broad H$\alpha$ emission-line profile 
\citep{1994ApJ...434...82G,1996PhDT........80M,1998ApJ...508..657M,1999MNRAS.303..227Y,2002MNRAS.335..773S}. 
Such features are naturally produced if the line emission originates in a rotating 
disk (presumably the outer regions of the accretion disk itself) and is scattered in a compact region that is co-planar with the disk, and closely surrounds the BLR 
\citep{2005MNRAS.359..846S}. 
This equatorial scattering region (ESR) is thus obscured by the torus in S2's and has the correct geometry to account for the observed alignment 
of $\theta_v$ with the radio position angle (RPA) in S1's.  

\begin{figure}
\plotone{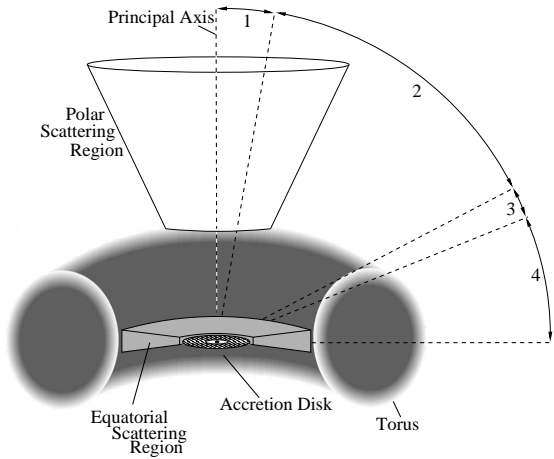}
\caption[The Refined AGN Unified Model]{
Refined AGN model of \cite{2004MNRAS.350..140S} showing the two-component scattering geometry proposed to explain the optical polarization characteristics 
of Seyfert nuclei. The ESR is modeled as a flared disk that closely surrounds the accretion disk. The PSR is modeled as a truncated cone aligned with the axis of the 
circum-nuclear torus. The two regions produce orthogonal polarization. We assume that the 
symmetry axes of the emission disk, both the scattering regions and the torus are co-aligned and define the principal axis of the system. The observed optical polarization
depends on orientation: if the line-of-sight lies in region 1 we would see a null-polarization 
S1, in region 2 an equatorially scattered S1, in region 3 a polar scattered  S1, and in region 4 an S2. 
}\label{fig:model}
\end{figure}

However, it is also evident that S1's as a class exhibit a much wider range of optical polarization properties than S2's 
\citep{2002MNRAS.335..773S, 2004MNRAS.350..140S}. In addition to the equatorially scattered objects, about 20\% exhibit null polarization, while $\sim25\%$ 
show characteristics of S2-like polar scattering. This diversity in the optical polarization properties of Seyfert nuclei can be understood if 
{\em both} polar and equatorial scattering regions are present in all Seyferts. The form of the observed polarization is then determined largely by the inclination 
of the torus axis to the line-of-sight \citep{2004MNRAS.350..140S}. See Figure~\ref{fig:model} for a schematic representation of this model. As inclination increases 
from pole-on to edge-on we first see null polarization S1's, then S1's dominated by equatorial scattering (where $\theta_v$ is aligned with the radio source axis). With 
a further increase in inclination we see polar-scattered S1's (with $\theta_v$ perpendicular to the radio axis), and finally S2's, like NGC 1068, that exhibit polarized 
broad-lines due to polar-scattering. 

In this model, therefore, the polar-scattered S1's occupy an orientation regime intermediate between unobscured S1's and a totally obscured S2's. The direct 
line-of-sight to the nucleus  passes through the torus at high latitudes, and is subject to a modest amount of extinction ($A_v\sim1-4$ magnitudes). 
The resulting 
attenuation of light from the ESR (within the torus) allows the orthogonally polarized flux from the extended polar scattering region 
(PSR; outside the torus) to dominate the net polarization.
In the context of
recent clumpy torus models \citep{2002ApJ...570L...9N,2006A&A...452..459H,2008ApJ...685..147N,2008ApJ...685..160N}, an S2 or a polar scattered S1 will be observed 
depending on whether or not the line of sight is intercepted by one or more dense clouds,  with the probability of this occurring being a decreasing function of 
inclination. Unlike in S2's, however, the relatively modest line-of-sight extinction in polar scattered S1's does not suppress the direct emission from the BLR. 

This simple model can in principle be tested by obtaining polarimetry observations at longer wavelengths. 
For example, at 2.0\micron\ extinction along direct lines-of-sight to the nucleus will be a factor 10 less than at $V$, suggesting that it may be possible to detect the 
polarization signature of the ESR at NIR wavelengths, even though polar scattering dominates at optical wavelengths.
Since the scattering geometry dictates that the polarization produced by the ESR must be orthogonal to that produced by the PSR, the presence of equatorial scattering is 
easily confirmed by comparing the polarization position angle 
at  2.0\micron\ ($\theta_{2\micron}$) with that measured in the optical ($\theta_v$). Furthermore, the compact ESR will only contribute to the polarization of the 
unresolved nucleus,
whereas the more extended PSR would be expected to dominated any off-nucleus polarization. 

Here we present NICMOS 2.0\micron\ imaging polarimetry observations of six polar-scattered S1's, obtained for this purpose. The high spatial resolution of 
{\it HST}/NICMOS (0\farcs2) 
is desirable as it minimizes cancellation between the two orthogonal polarization states and may perhaps even resolve the two scattering regions.

The paper is organized as follows. In \S~\ref{obs} we outline the sample selection and describe our observations and data reduction. 
Our approach to the polarimetric analysis of the data is described in \S~\ref{ptheta} and the results are presented in \S~\ref{results}.   
We discuss these results in the context of the two component scattering model in \S~\ref{discussion} and present our conclusions in \S~\ref{cons}. 

\section{Observations and Data Reduction}\label{obs}

The six S1's selected for this study were drawn from the list identified by  \citet{2004MNRAS.350..140S} as showing clear spectropolarimetric signatures of polar 
scattering. In general, polar scattered S1's  are characterized by a systematic increase in the degree of polarization over the optical spectrum, while the polarization 
position angle is approximately constant, without the large excursions over the broad Balmer lines that are commonly seen in equatorially scattered S1's. For the objects 
studied here, the range in $p_v$  over the wavelength range $\sim$4500 -- 7100\AA\ and the value of $\theta_v$, averaged over the same range, are listed for each object  
in Table~\ref{tab:sample}, along with several other properties of interest.  

\begin{deluxetable}{lccc}
\tablecaption{Details of the Observations\label{tab:acc}}
\tablewidth{0pt}
\tablehead{
\colhead{Target}&
\colhead{Orientation (\degr)}&
\colhead{S/N}&
\colhead{$p$}
}
\startdata
Fairall 51           & 33.4 & 281 & 1.0   \\
NGC 4593         & 67.0 & 193 & 1.4 \\
NGC 3227         & 75.6 & 106 & 1.9  \\
Mrk 766            & 82.6 &   57 & 2.7  \\
Mrk 1239          & 53.2 & 422 & 1.0  \\ 
ESO 323-G077 & 77.0 & 525 & 0.8 \\
\enddata
\tablecomments{
Frame orientations for the sample (given by the NICMOS ``ORIENTAT'' header parameter) and the theoretically detectable polarization ($p$) for the measured S/N.
The S/N was determined from an aperture radius of 7.6 pixels (0\farcs58), corresponding to the unresolved nucleus. The values of $p$ are the minimum values for which a 
$3\sigma$ detection is possible.
}
\end{deluxetable}

The observations were designed to mitigate or facilitate removal of several known NICMOS camera 2 (NIC2) defects such as residual bad pixels, latent image persistence, 
and the inter-pixel response functions (IPRFs). A three point spiral dither pattern, with a point spacing of 3\farcs0 (40 pixels), was executed a total of four times 
through each polarizer 
using an exposure time of 10 seconds (a total of 12 pointings per polarizer). The patterns were offset from each other by 5\farcs0. These 
observations sample the nucleus of each target and minimize the possibility of persistence, thoroughly sample the IPRF and allow the detection 
of residual bad pixels. To sample the background, 10 second images were taken through each polarizer after a 60\farcs0 offset. To check for latent 
image persistence, we performed aperture photometry on individual reads in the {\tt multiaccum} images. Curves of growth (count rate {\it vs.} 
time) were constructed and found to be linear over the entire exposure. In addition, the residuals between image sections both before and after 
illumination by the bright point sources were checked, and found to be consistent with the generic noise characteristics of NIC2. Therefore, we find 
no evidence of the latent images expected to result from persistence. 

\begin{figure}
\plotone{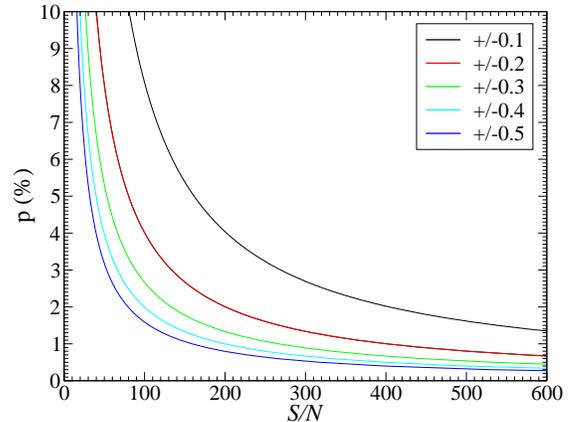}
\caption[Results from Aperture Photometry]{
Theoretically achievable polarization measurements as a function of S/N. The curves correspond to increasing uncertainties in $p$ (given in the 
inset). Derived from \cite{1999PASP..111.1298S}.
}\label{fig:acc}
\end{figure}

All data were first passed through the {\tt calnica} pipeline with all calibration switches turned on. Accurate 
polarimetric analysis requires that each individual pointing be treated separately. Therefore these data were not processed through the 
{\tt calnicb} pipeline. The sky images were used to subtract the background and residual dark current. As the dithered exposures have not been 
processed as an association ({\tt calnicb}), they still suffered from residual bad pixels and ``grot'', i.e., pixels with reduced throughput due to particulate 
contamination on the detector. In these cases we used the data quality image extensions, and the default ``mask-file'' to create a bad pixel mask that 
was a combination of hot and cold pixels, and pixels affected by grot. These pixels were then repaired using a 2D interpolation 
from the surrounding unaffected pixels. 

\section{Polarimetric Analysis}\label{ptheta}

The use of NICMOS as an imaging polarimeter has been discussed previously by several authors \citep[e.g.,][]{2000PASP..112..983H,2006PASP..118..642B,
2009PASP..121..153B}. Our polarization measurements were made using the procedures and calibrations described in \cite{2009PASP..121..153B}. 
Radial profiles in the Stokes parameters Q and U were constructed for each individual pointing using the 
{\tt digiphot} package within IRAF\footnote{IRAF is distributed by the National Optical Astronomy Observatories, which are operated by the 
Association of Universities for Research in Astronomy, Inc., under cooperative agreement with the National Science Foundation.}. 
The average per-pixel radial profiles were then computed using an iterative $2\sigma$ clipping procedure. 
In general, rejected profiles were affected by  uncorrectable bad pixels and ``grot'', and typically amount to 10\% of the data for each source.
However, in the case of ESO 323-G077 three of the 12 profiles had to be removed. The $1\sigma$ dispersion around the average for the pointings 
is used to define the statistical uncertainty at each radial point.

\begin{deluxetable*}{lcccccc}
\tablecaption{The $2.0\mu{\rm m}$ Polarization Properties of the Sample \label{tab:results}}
\tablewidth{0pt}
\tablehead{
\colhead{Target}&
\colhead{$p_{2\micron}(\%)$}&
\colhead{$\theta_{2\micron}(\degr)$}&
\colhead{$\delta{\theta}(\degr)$}&
\colhead{$p_{2\micron}(\%)$}&
\colhead{$\theta_{2\micron}(\degr)$}&
\colhead{$\delta{\theta}(\degr)$}\\
\colhead{}&
\colhead{(0\farcs58)}&
\colhead{(0\farcs58)}&
\colhead{(0\farcs58)}&
\colhead{(1\farcs5)}&
\colhead{(1\farcs5)}&
\colhead{(1\farcs5)}\\
}
\startdata
Fairall 51           & $2.2\pm0.4$ & $139\pm8$ & $1\pm8$   & $1.9\pm0.4$ & $148\pm8$ & $8\pm8$    \\ 
NGC 4593         & $0.4\pm0.6$ & \nodata       & \nodata    & $1.5\pm0.4$ & $77\pm6$   & $68\pm6$  \\
NGC 3227         & $0.5\pm0.5$ & \nodata       & \nodata    & $0.7\pm0.4$ & $102\pm20$ & $23\pm20$ \\         
Mrk 766            & $0.6\pm0.7$ & \nodata       & \nodata    & $2.7\pm0.6$ & $43\pm5$   & $47\pm5$   \\
Mrk 1239          & $0.7\pm0.4$ & $83\pm9$   & $46\pm9$ & $0.9\pm0.3$ & $63\pm8$   & $66\pm8$ \\
ESO 323-G077 & $1.0\pm0.4$ & $95\pm10$ & $11\pm10$ & $0.9\pm0.3$ & $74\pm8$   & $10\pm8$  \\
\enddata
\tablecomments{
Values of $p_{2\micron}$ and $\theta_{2\micron}$ for the sample. Values of $\theta_{2\micron}$ and 
$\delta\theta$ are taken from apertures of radius 0\farcs58 and 1\farcs5 and are given in degrees in the celestial (on sky) reference frame. The values 
of $\delta\theta$ give the differences between the NIR and the average optical polarization position angle (Table~\ref{tab:sample}). No significant polarization
was measured within the nuclear apertures of NGC\,4593, NGC\,3227 and Mrk\,766.
}
\end{deluxetable*}

\begin{deluxetable}{lccccc}
\tablecaption{Annular Polarimetry Results \label{tab:annres}}
\tablewidth{0pt}
\tablehead{
\colhead{Target}&
\colhead{$p_a(\%)$}&
\colhead{$\theta_a(\degr)$}&
\colhead{$\delta{\theta_{0\farcs58}}(\degr)$}&
\colhead{$\delta{\theta_{1\farcs5}}(\degr)$}\\
}
\startdata
Fairall 51          & $0.2\pm0.4$ & \nodata & \nodata  & \nodata    \\
NGC 4593        & $0.5\pm0.3$ & $41\pm9$ & \nodata  & $36\pm15$ \\
NGC 3227        & $0.4\pm0.1$ & $166\pm7$ & \nodata  & $44\pm27 $    \\
Mrk 766            & $1.6\pm0.4$ & $24\pm9$ & \nodata  & $19\pm14$ \\
Mrk 1239          & $1.5\pm0.2$ & $144\pm6$ & $61\pm15$ & $81\pm14$\\
ESO 323-G077 & $0.9\pm0.2$ & $83\pm8$ & $12\pm18$ & $9\pm16$  \\
\enddata
\tablecomments{
Percentage polarization ($p_a$) and polarization position angle $\theta_a$ at 2\micron\ 
for a circum-nuclear annulus extending from 0\farcs58 to 1\farcs5. The last two columns give the difference in $\theta_{2\micron}$ between
the annular value ($\theta_a$) and the aperture values for radii of 0\farcs58 and 1\farcs5, respectively.  
}
\end{deluxetable}

Using the recipe of \cite{1999PASP..111.1298S}, the final Stokes parameter profiles were combined to determine the percentage polarization, position angle
and associated uncertainties as functions of aperture radius. The frame orientations of the observations, which must be subtracted from the calculated polarization 
position angle in order to determine $\theta_{2\micron}$ in the celestial reference frame,  are given in Table~\ref{tab:acc} (column 1). We define the unresolved nucleus 
to be the
area enclosed by the  $2^{\rm nd}$ Airy maximum of the {\it HST}+NICMOS PSF, which corresponds to 7.6 pixels or 0\farcs58. The second column of Table~\ref{tab:acc} gives 
the integrated S/N within this nuclear aperture for each object. Following \citeauthor{1999PASP..111.1298S}, we have computed the detectable polarization and its 
associated uncertainty as a function of S/N. The polarization detectable to a given uncertainties is shown for several values of the latter in Figure~\ref{fig:acc}. 
The limiting polarization detectable at the $3\sigma$ level 
for the S/N within the nuclear apertures of our objects  is listed in the third column of Table~\ref{fig:acc}. 

\begin{figure*}
\plotone{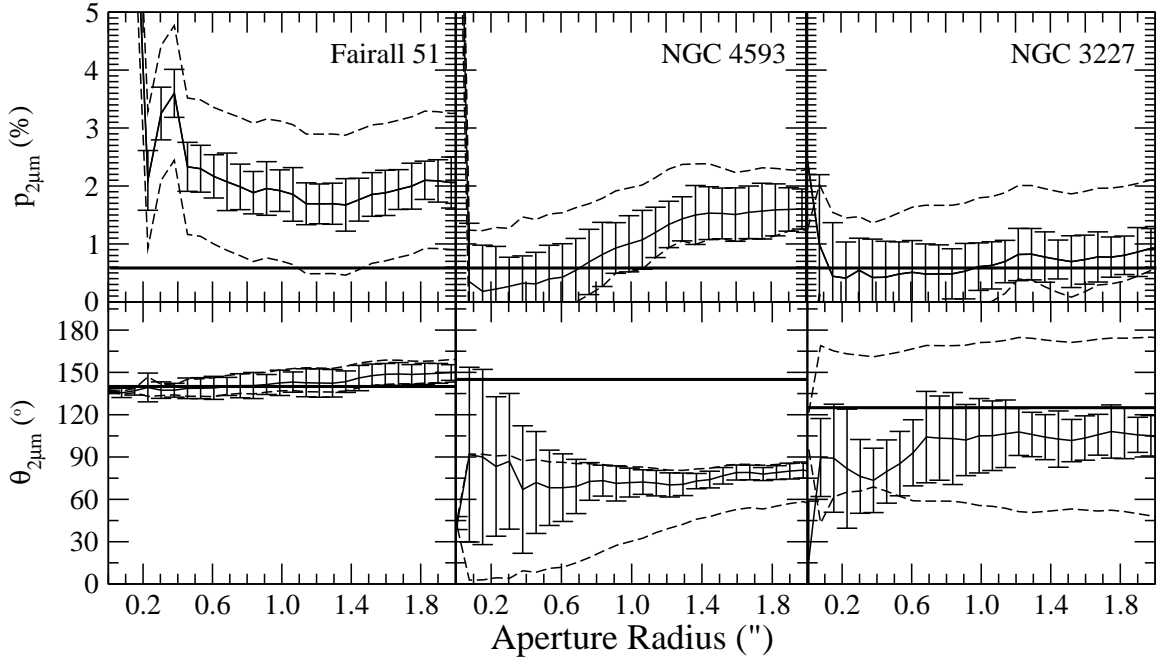}
\caption[Results from Aperture Photometry]{
Radial polarization profiles at 2.0\micron, derived from aperture polarimetry, for Fairall 51, NGC 4593 and NGC 3227. At each point, the polarization is determined  
from the enclosed Stokes fluxes. The error bars show the statistical uncertainties arising from photometric repeatability. The dashed lines show the possible range in 
values 
corresponding to the systematic uncertainty on the parallel transmission coefficients. {\it [Top Row:]} Radial percentage polarizations at 2.0\micron. The solid thick 
line at $p_{2\micron} = 0.6\%$ is the upper limit to the instrumental polarization. {\it [Bottom Row:]} 2.0\micron\ position angles. The thick solid lines mark the 
optical polarization position angles ($\theta_v$).}\label{fig:nuc1}
\end{figure*}

\begin{figure*}
\plotone{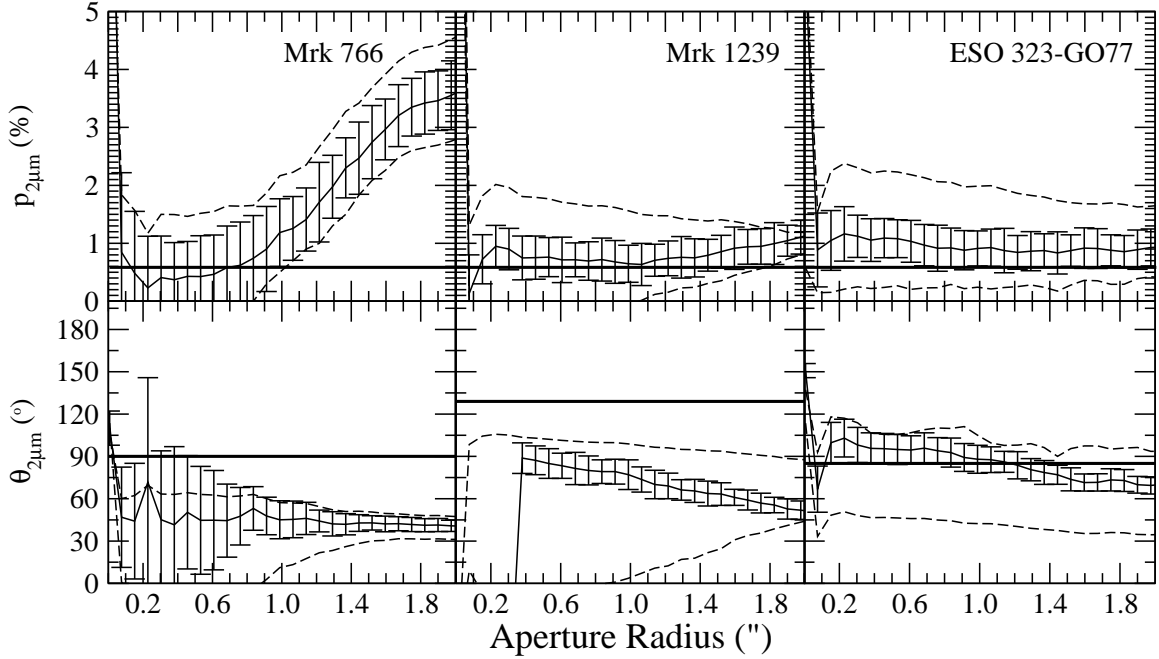}
\caption[Results from Aperture Photometry]{
As Fig~\ref{fig:nuc1} but for Mrk 766, Mrk 1239 and ESO 323-G077. 
}\label{fig:nuc2}
\end{figure*}

The radial profiles in $p_{2\micron}$and  $\theta_{2\micron}$ for each member of the sample are presented in Figures~\ref{fig:nuc1} and \ref{fig:nuc2}. On these diagrams, 
the solid bars indicate the statistical uncertainty. We note that the statistical errors are correlated, as $p_{2\micron}$and  $\theta_{2\micron}$ are determined from radial
integrals of the Stokes fluxes. The 
$p_{2\micron}$ panels also indicate the {\em upper limit} of 0.6\% to the instrumental polarization of NICMOS 2, as determined by \cite{2009PASP..121..153B}. The average
optical polarization position angle (Table~\ref{tab:sample}) is shown in the $\theta_{2\micron}$ panels. The NICMOS polarimetry is also subject to  a systematic error,
associated with the uncertainties in the parallel transmission coefficients $t_k$, which will be the same for all objects. While this can shift the values of both 
$p_{2\micron}$ and $\theta_{2\micron}$, the radial trends will not be affected. The dashed lines in  Figures~\ref{fig:nuc1} and \ref{fig:nuc2} indicate the allowed ranges 
of these shifts.
Table~\ref{tab:results} lists the values $p_{2\micron}$ and $\theta_{2\micron}$ corresponding  to aperture of radii of 0\farcs58 and 1\farcs5. 

The radial profiles contain information on the spatial variation of polarization. However, any extended source of polarization will be contaminated by the signal from the
unresolved nucleus. Therefore, we have also used the techniques outlined above to determine the polarization within an annular aperture extending to 1\farcs5, but 
excluding the  
inner 0\farcs58 of each target. The resulting values of  $p_{2\micron}$ and $\theta_{2\micron}$ are given in Table~\ref{tab:annres}.

\section{Results}\label{results}

We have detected significant polarization in all six objects, although the radial profiles exhibit diverse behavior. At first sight, the  radial  profiles 
derived from aperture polarimetry suggest that in three objects (NGC\,3227, Mrk\,1239 and ESO\,323-G077), $p_{2\micron}$ is consistent with the upper bound
on the instrumental polarization. However, the measured values of $\theta_{2\micron}$ differ significantly from object to object in the instrumental plane, indicating
that the measured polarization cannot be solely instrumental in origin in all three objects. Furthermore, Mrk\,1239 exhibits polarization significantly higher than the 
instrumental upper limit in the off-nuclear annulus, as will be discussed in more detail below. Two of our sample, Fairall\,51 and Mrk\,1239, are also found in NIR 
polarization study
of \cite{1990MNRAS.244..577B}. For Fairall\,51, our results are consistent with \citeauthor{1990MNRAS.244..577B}'s J \& H photo-polarimetry, which was obtained from a
 6\farcs0 aperture. The case of  Mrk\,1239 is discussed below. 
 
Only one source, Fairall 51, exhibits radial polarization profiles consistent with a point source. It's $p_{2\micron}$ and  $\theta_{2\micron}$ profiles are similar to 
those presented by \cite{2009PASP..121..153B} for polarized standard stars, which typically have large fluctuations at small radii, due to low S/N, but stabilize at 
constant values of $p_{2\micron}$ and  $\theta_{2\micron}$ beyond the third Airy ring. In addition, no significant polarization was detected in the off-nuclear annulus 
for this object (Table~\ref{tab:annres}). The value of $\theta_{2\micron}$ derived from aperture
polarimetry is the same, to within 1$\sigma$, as the average optical polarization position angle (Tables~\ref{tab:sample} and~\ref{tab:results}).

All 5 of the remaining objects show evidence for extended sources of polarization. In fact, the unresolved nucleus is not significantly polarized in 3 of these objects -- 
NGC\,4593, NGC\,3227 and Mrk\,766. In NGC\,4593 and Mrk\,766, the $p_{2\micron}$ profiles show a strong increase with aperture radius once this exceeds 0\farcs58, 
resulting in significant polarizations of, respectively,
$1.5\pm0.4$\% and $2.7\pm0.6$\% within the 1\farcs5 aperture (Table~\ref{tab:results}). Both objects also exhibit polarization in the off-nuclear annulus (although only at 
the $\approx 2\sigma$ level in NGC\,4593). However, in both cases  the values of  $\theta_{2\micron}$, whether measured from the 1\farcs5 aperture or the annulus, differ 
widely from the optical polarization position angle. NGC\,3227 yields only a marginal ($\approx 2\sigma$) detection of  $p_{2\micron}$ in the 1\farcs5 aperture. However, 
this object is significantly, albeit weakly, polarized in the off-nuclear annulus. The values of $\theta_{2\micron}$ measured from the aperture and annulus bracket the 
average optical value (Tables~\ref{tab:results} and ~\ref{tab:annres}). 

ESO\,323-G077 is polarized at the level of $\approx 1$\% in both the unresolved nucleus (a $2.5\sigma$ detection) and in the off-nuclear annulus. The values of 
$\theta_{2\micron}$ measured from the two regions are consistent with each other and also with the average optical position angle.

Finally, Mrk\,1239 is unique among the objects studied here in that it exhibits strong evidence for a change in polarization state between the nucleus and
the surrounding annulus. 
The unresolved nucleus is weakly polarized; our data yield only a $\approx 2\sigma$ detection. However, the radial profile
in $p_{2\micron}$ shows evidence for a systematic increase to $\approx 1$\% beyond 1\farcs4. $\theta_{2\micron}$ also varies systematically with radius, 
decreasing from $\approx 80\degr$ in the unresolved nucleus to $\approx 60\degr$ for radii $\gtrsim 1\farcs5$. The orientation of the polarization 
vector at 2\micron\ therefore differs by $\approx 50-70\degr$ from the average optical value.  

In the 0\farcs58 -- 1\farcs5 annulus, however, the polarization is significantly higher, at $1.5\pm0.2$\%, than in the nucleus and also 
quite closely aligned (within 15\degr) with the average optical polarization (Tables~\ref{tab:sample} and ~\ref{tab:annres}).  Mrk\,1239 is the only object in the sample,
for which the annular polarization has a position angle significantly different to those measured in either the 0\farcs58 or 1\farcs5 apertures (Table~\ref{tab:annres}). 
Perhaps
more significant is the comparison with ESO\,323-G077, which is the only other object in which we have detected polarization in {\em both} the unresolved nucleus and the 
annulus.
In this object, the 2\micron\ polarization in both regions has the same position angle as the optical spectrum. In Mrk\,1239, on the other hand,
there is a $\ge 60\degr$ change in position angle between the nucleus and the circum-nuclear annulus, with the  latter having approximately the 
same position angle as measured in the optical. 

The systematic change in $\theta_{2\micron}$ with increasing aperture radius (Fig.~\ref{fig:nuc2}) can be understood as the result of mixing of the Stokes fluxes 
from the unresolved nucleus and the surrounding extended source of polarization implied by the annular polarimetry. As the aperture radius increases, and encompasses
the PSF, the contribution to the integrated Stokes fluxes from the extended source increases relative to that from the nucleus, causing a gradual change in 
$\theta_{2\micron}$
to angles intermediate between the intrinsic values for the two sources. In fact, in Fig.~\ref{fig:nuc2}, $\theta_{2\micron}$ changes towards the direction of the 
supplementary angle of $\theta_a$.

Our annular polarization measurements for Mrk\,1239 are consistent with the J \& H photo-polarimetry of \cite{1990MNRAS.244..577B}. This suggests that the
extended source, not the nucleus, dominates the polarization measured within the 6\farcs0 aperture used by these authors. 

\section{Discussion}\label{discussion}

\subsection{Predictions of the unified scattering model}\label{predictions}

The polar scattered S1's studied here are remarkable in that while their total flux optical spectra 
are typical of the general population of S1's, their polarization spectra show characteristics similar to those of  ``hidden broad-line'' S2's. As previously outlined, 
these 
properties can be understood if polar-scattered S1's occupy an orientation regime (Fig.~\ref{fig:model}) in which the direct line-of-sight to the nucleus 
follows a relatively transparent path through the torus, and is thus subject to modest visual extinction ($A_V\sim1-4$ mag). 

In our NICMOS images, the compact ESR should contribute to the polarization only within the unresolved nucleus, whereas the PSR 
should dominate off-nucleus, if it is resolved. If the ESR is the dominant source of  polarized flux at 2\micron, the nuclear point source will have a polarization 
position angle perpendicular to that obtained from the optical spectropolarimetry. Conversely, we would expect any circum-nuclear polarized flux to come from the PSR 
and hence have a polarization vector aligned with that of the optical polarization. 

\subsection{Has the equatorial scattering region been detected?}\label{equatorial}

Our 6 targets exhibit a range of 2\micron\ polarization properties, indicating that the origin of the NIR polarization is more complex than envisaged in the basic 
two-component 
scattering model. 

\subsubsection{Mrk~1239: equatorial and polar scattering}
This object presents the most compelling case for the detection of an ESR. The simplest interpretation of the annular polarimetry is that the ESR
dominates the polarization of the unresolved nucleus, whereas at radii $>$0\farcs6, where the influence of the nuclear PSF is negligible, the 
polarization is dominated by the PSR. The clear detection of the ESR signature within 0\farcs6 implies that the size of this region is less than 
$\sim 200$\,pc, whereas the annular aperture within which the PSR dominates the measured polarization corresponds to a linear scale $\sim 200-600$\,pc. 
However, it should be noted that the inner radius of the annulus was chosen to exclude the nuclear PSF; there is no reason to believe it represents a 
physical boundary between the two regions. Indeed, we expect the ESR to be located within the torus, with a size limited by the dust sublimation radius 
($\sim 0.05$\,pc for Mrk\,1239; see Section~\ref{scatgeom}). It is also likely that the PSR extends much further inwards than 200\,pc, but contributes 
a smaller fraction of the combined stokes fluxes within the nuclear aperture than the ESR.

Mrk\,1239 is a Narrow Line Seyfert 1 galaxy \citep{1989ApJ...342..224G} that has a number of remarkable properties.  Its optical polarization is 
among the highest measured in a S1, and spectropolarimetry shows that the broad Balmer lines are redshifted in polarized flux. This indicates that 
the polar-scattering medium is undergoing outflow in a  $\sim 1000$km\,s$^{-1}$ wind (\citealt{1989ApJ...342..224G}; Robinson et al., 2010a, in 
preparation). It also has an exceptionally steep soft X-ray spectrum, which exhibits warm absorbers \citep{2004AJ....127.3161G},  and a strong NIR 
excess peaking at $\sim2$\micron\ \citep{2006MNRAS.367L..57R}. A comparison of spectral energy distributions  leads \citeauthor{2006MNRAS.367L..57R} 
to infer that the AGN in Mrk\,1239 is subject to an intrinsic reddening $E(B-V) = 0.54$, corresponding to a visual extinction $A_V\sim 1.7$ mag. 
This is consistent with the \cite{2004MNRAS.350..140S} estimate of the extinction along the direct path to the nucleus in polar-scattered S1's, and 
therefore it is entirely plausible that we are ``seeing through'' to the ESR in the NIR. The NIR excess can be attributed to black body emission from 
hot dust at a temperature $\sim 1200$\,K \citep{2006MNRAS.367L..57R}. 

\subsubsection{NGC\,4593 and Mrk\,766}
Two other objects, NGC\,4593 and Mrk\,766, also exhibit large differences ($\delta\theta > 60\degr$) between the $\theta_v$ and $\theta_{2\micron}$  PA's. However, in both 
cases the polarization of the 
unresolved nucleus (within $\sim$0\farcs6) is consistent with zero, the integrated polarization only becoming significant at larger radii. The measured values of 
$p_{2\micron}$ and $\theta_{2\micron}$ must therefore be associated with an extended scattering region. In these objects, therefore, large values of $\delta\theta$ do not 
imply the 
detection of the compact ESR. In fact, we would expect  
off-nuclear polarization to be due to polar scattering and hence parallel to the optical polarization.  What then does account for the different polarization position 
angles in the optical and NIR? The measured polarization position angle in 
either wave-band could be influenced by other sources of polarization. An obvious source is dichroism due to aligned dust grains either in the host galaxy, or our own. 
This is likely to be particularly significant in NGC\,4593, which has a relatively low optical polarization (0.5\%). For example, \citet{2005MNRAS.363.1241H} estimate 
that the optical polarization angle could be uncertain by up to $90\degr$ due to Galactic interstellar polarization (ISP) alone. Thus, we have to regard the comparison 
between the 2\micron\ and optical polarization PA's as highly uncertain in this case. 

On the other hand, ISP contamination cannot similarly explain the observed large $\delta\theta$ observed in Mrk\,766, which has much higher optical polarization (2--4\%). 
\citeauthor{2005MNRAS.363.1241H} estimate that Galactic ISP would change the optical polarization PA  by $<5\degr$. In this case, therefore, we must conclude that 
polarized flux arising from a different scattering geometry makes a 
substantial contribution to the off-nucleus NIR polarization. The nature of this scattering geometry is unclear, except that the average scattering plane must be 
$\sim 45\degr$ different
from that of the region responsible for producing the optical polarization.

\subsubsection{Polar scattering}

In Fairall 51 and ESO\,323-G077 the 2\micron\ polarization is closely aligned with the optical position angle, indicating that polar scattering dominates in the NIR. The 
PSR appears to be unresolved in Fairall 51, implying a linear size scale $< 150$\,pc. A simple luminosity scaling of the PSR size inferred for Mrk\,1239 ($r\propto L^{1/2}$) 
predicts that this region should be resolved in Fairall 51; the fact that it is not suggests a more compact distribution of scatterers in this object. Significant circum-nuclear 
polarization is detected in ESO\,323-G077, indicating that the polar scattering region is resolved. Evidence for extended polarization is also detected in NGC\,3227, and we 
similarly attribute this to polar scattering, although the value of $p_{2\micron}$ is relatively low and $\theta_{2\micron}$ is not as clearly aligned with $\theta_v$. 
The low annular polarization is probably the result of geometrical cancellation of the polarization vectors. 

\subsection{The complex 2\micron\ polarization properties of polar-scattered Seyfert 1's}\label{properties}

Despite having similar optical polarization spectra, our  small sample of 6 objects exhibits 3 different kinds of behavior at 2\micron. 
In this section, we discuss some possible  causes of the complexity evident in our results.

\subsubsection{Spatial resolution}\label{resolution}

Within the unresolved nucleus it is possible that both 
the equatorial and polar scattering regions contribute to the polarized flux at 2\micron. The Stokes fluxes from these two sources will tend to cancel; if for example, 
the polarizations are precisely orthogonal, the result will be a net polarization either parallel or perpendicular to the system axis, depending on which region produces 
the higher stokes flux. The net polarized flux will be lower than would be the case if only one of the scattering regions contributes. In principle, this effect could 
account for the weak polarization in the unresolved nuclei of Mrk\,766, NGC\,4593 and NGC\,3227. Similarly, in Fairall\,51 and ESO\,323-G077,  the fact that the PSR 
appears to dominate the 2\micron\ polarization 
does not necessarily imply that the ESR is not present in these objects, simply that it produces relatively smaller Stokes fluxes at 2\micron. 

If spatial resolution is a key factor in determining the observed polarization, we might expect the 2\micron\ polarization properties to be related to the linear scale 
(listed in Table~\ref{tab:sample}). However, there is no evidence that this is the case, 
suggesting that the spatial resolution of the observations (and consequent mixing of orthogonal polarization states) is not the sole cause of the range of observed 
behavior.  
Indeed Mrk\,1239,  in which the two scattering regions are resolved, is the most distant object in our sample. In reality, the relative contributions of the ESR and PSR 
to the net polarization in the unresolved nucleus will be determined by multiple variables including the geometry and density of the scattering medium, the physical size 
of the PSR and,
as we discuss below, the line of sight extinction.

\subsubsection{Extinction}\label{extinction}

If the amount of dust extinction along the direct line-of-sight to the nucleus is the dominant effect in determining whether or not the polarization signature of the ESR 
is actually observed in a given object, we might expect a correlation between nuclear reddening and the observed 2\micron\ polarization properties. Unfortunately, 
determining the nuclear reddening in AGN is problematic \citep{1983ApJ...268..591G}. The most 
accessible nebular diagnostic, the Balmer decrement, does not have a well determined intrinsic value for the physical conditions pertaining in the broad line region. 
For example, cloud ensemble photoionization models computed by \citet{2004ApJ...606..749K}, indicate 
H$\alpha$/H$\beta$ ratios in the range 3.7--4.9. On the other hand, recent empirical studies of the broad-line Balmer decrement in quasars indicate a mean value only 
slightly greater than the Case B recombination value \citep{2008MNRAS.383..581D}. Within our sample, there is a fairly wide range in the broad-line H$\alpha$/H$\beta$ 
ratio (3.5 to 6.1, Table~\ref{tab:sample}), but there is no correlation with $\delta\theta$. Moreover, the uncertainties are such that there is no significant difference in 
H$\alpha$/H$\beta$ between Mrk\,1239, the object which (based on our results) we would expect to be least affected by reddening, and either Fairall 51 or ESO\,323-G077, 
the objects we would expect to be most affected. If extinction differences are indeed responsible for the observed range in polarization behavior, then these variations 
are masked by 
the uncertainties in the measurement of the Balmer decrements, or by variations in the intrinsic broad-line Balmer decrements.

\subsubsection{Scattering geometry}
\label{scatgeom}

We must also consider the possibility that the effective scattering geometry for 2\micron\ radiation differs from that seen by optical photons. 
In the two component scattering model, the ESR is 
located within the torus, and has a flattened, annular structure co-axial with the accretion disk. In order to produce the characteristic features of the H$\alpha$ 
polarization, 
notably $\theta(\lambda)$ variations across the line, it must also closely surround the H$\alpha$ emitting region of the disk \citep{2005MNRAS.359..846S}. The  
NIR emission from AGN is believed to be dominated by thermal  radiation from dust heated by the optical-UV continuum of the accretion disk;  the 2\micron\ continuum 
coming predominantly from the hottest dust near the inner edge of the torus. It is unclear if this dust is part of the torus itself, or whether it is a distinct component.
\citet{2009ApJ...705..298M} find that a hot black body component ($T\sim 1400\,K$), in addition to a clumpy torus, is required to fit  the IR spectral energy distributions 
of low redshift quasars, with the black body component dominating at wavelengths $\lesssim 4\micron$.  Similarly, hot dust components have been found in several Seyfert 
galaxies, including Mrk\,1239 (as already noted), Mrk\,766 and NGC\,4593 (see Table 1 in \citealt{2009ApJ...698.1767R}). 

Whether or not it is part of the torus, the emitting region must be relatively compact; the radius at which dust grains have an equilibrium temperature 
$T\sim 1500$\,K is $r_{d} \sim 0.4L_{45}^{0.5}$\,pc, where $L_{45}$ is the optical-UV luminosity in units of $10^{45}$\,erg\,s$^{-1}$ (\citealt{2008ApJ...685..147N}; 
see also \citealt{1987ApJ...320..537B}). If we take values of  $\lambda L_{\rm5100\AA}$, measured from our 
spectropolarimetric data, to be characteristic of the optical-UV luminosity of our low-to-moderate luminosity Seyferts, then $L_{45}\sim 3\times 10^{-3}- 2\times 10^{-2}$ 
and hence $r_{d}\sim 30-60$\,ld. 

The BLR lies within the torus but also scales in size approximately as $L^{0.5}$ 
\citep{2005ApJ...629...61K, 2009ApJ...697..160B}. Using \citeauthor{2009ApJ...697..160B}'s calibration of the BLR radius -- optical luminosity scaling relation, we find 
$r_{BLR}\sim 10$\,ld for our objects. 

\begin{figure}
\plotone{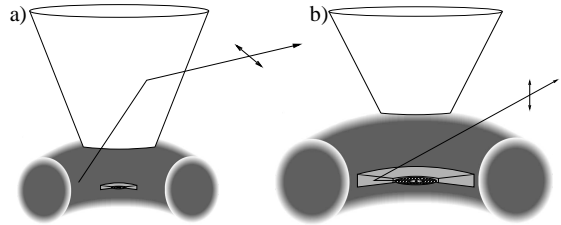}
\caption[Possible Models]{(a) Polar scattering of NIR emissions from the inner edge of the torus. The solid line marks the light path. The solid line with double arrows 
shows the position angle of the polarization vector. In this case it is perpendicular to the principal axis due to polar scattering. Here, the ESR is a factor $\sim 5$ 
smaller than 
the NIR emitting region.  (b) Equatorially scattered 2\micron\ emission from the outer edge of the accretion disk. In this case, the 
position 
angle of the polarization vector is aligned with the principal axis, hence equatorial scattering is detected. 
}\label{fig:model1}
\end{figure}

These calculations suggest, therefore, that the 2\micron\ dust-emitting region is typically a factor 
$\sim 5$ larger than the BLR, and hence also the ESR, if the latter closely surrounds the BLR\footnote{Note, however, that in a NIR reverberation mapping study of 4 nearby 
Seyferts, including NGC\,3227,  \citet{2006ApJ...639...46S} find that the NIR emission region is only
a factor $\sim 2$ larger than the BLR.}. 
The ESR may therefore be much smaller than the hot dust NIR emission source and will intercept and scatter only 
a small fraction of the emitted flux. On the other hand, the PSR must extend beyond the torus and must be at least 
comparable in size with and probably much larger than the NIR-emitting inner region. Therefore, polarization of 2\micron\ emission due to hot dust located in or near the 
inner region 
of the torus is likely to be dominated by {\em polar scattering} (Fig.~\ref{fig:model1}a).

However, the outer regions of the accretion disk are also expected to emit NIR radiation. Recently, \citet{2008Natur.454..492K} reported that in several quasars, the 
NIR continuum in polarized light exhibits an $f_\nu \propto \nu^{1/3}$ spectrum, consistent with the characteristic long wavelength spectrum of the accretion disk. 
It appears that in these objects, it is the accretion disk emission, not the torus dust emission that is polarized. As 
the quasars they studied do not exhibit broad line polarization, \citeauthor{2008Natur.454..492K} argue that the scattering region is located well outside the NIR emitting 
region of 
the accretion disk but {\em within} the radius of the BLR. A simple black body accretion disk model indicates that the disk will be cool enough ($\sim 1500$\,K) to emit at 
2\micron\ at a radius $r_{2\micron}\sim 7\times10^3R_g(L/L_{Edd})^{1/3}(M_\bullet/10^7M_\odot)^{-1/3}$, where $L_{Edd}$ is the Eddington luminosity, 
$R_g=2GM_\bullet/c^2$, and we have assumed a radiative efficiency of 10\%. For the lower luminosity Seyferts studied here, black hole masses estimated either by 
reverberation mapping \citep{2004ApJ...613..682P, 2009ApJ...705..199B}, or by  the 
virial method (e.g., \citealt{2006ApJ...641..689V, 2005ApJ...630..122G}, see also \citealt{2004ApJ...613..682P}) lie in the range  $M_\bullet\sim 0.2-4\times 10^7M_\odot$. 
Therefore, given that $L/L_{Edd} \le 1$, the simple disk model suggests that the 2\micron\ emission comes from radii $\lesssim 10$\,ld, a region comparable in size with 
Balmer line 
emitting BLR. In general, therefore, the NIR emitting region of the disk lies close to but within the ESR and accordingly we expect that 2\micron\ {\em accretion disk} 
emission should 
be mainly  subject to equatorial scattering, producing polarization parallel to the disk axis (and hence perpendicular to the observed optical polarization, 
Fig.~\ref{fig:model1}b).

These arguments suggest that the observed 2\micron\ polarization may in general depend on the relative importance of scattering of torus emission by the PSR, 
or scattering of accretion disk emission by the ESR. As our objects are, by selection, dominated by polar scattering in the optical, and as dust emission dominates  
the AGN continuum at wavelengths $\ge 2$\micron, it might be expected that polar scattering will dominate the 2\micron\ polarization even if the the direct line of sight 
to the 
ESR is essentially transparent. This appears to be the case in Fairall 51 and ESO\,323-G077. On the other hand, the two scattering regions are apparently resolved in
Mrk\,1239. We can plausibly attribute the polarization of the unresolved nucleus to equatorial scattering of the accretion disk emission, whereas the polarization detected 
off-nucleus
is due to polar scattering of the hot dust emission. Thus, while the latter component might dominate the total 2\micron\ flux of the unresolved nucleus, it does not 
contribute to the polarization {\em in the nucleus}.

The inner walls of the torus itself may also scatter 2\micron\ emission from the disk and/or hot 
dust components. Scattering of light from a central point source by the inner walls of an optically 
thick torus can produce polarization either parallel to or perpendicular to the axis, depending on 
the torus opening angle and inclination \citep{1995ApJ...452..565K,2000MNRAS.319..685C,2007A&A...465..129G}
As the dust sublimation radius is unresolved, this polarization would only contribute to the net 
polarization of the unresolved nucleus. However, such a contribution is unlikely to be significant. 
If the NIR emission is dominated by hot dust in or near the torus opening, i.e., approximately 
co-spatial with the scattering region, there will be a wide range in scattering angles leading to 
cancellation of the Stokes fluxes and a low net polarization. On the other hand, the monte-carlo 
simulations of \cite{2007A&A...465..129G} indicate that the polarized flux produced by torus scattering 
of disk emission, i.e., approximately a central point source, will be much smaller than the polarized 
fluxes produced by the PSR or ESR, for any plausible electron scattering optical depth.

\subsection{Wavelength dependence of polarization degree}\label{wavdep}

A key property of the polar scattered Seyferts is that the optical continuum polarization increases strongly to the blue. This trend continues into the NIR, in the 
sense that the  2\micron\  polarization of the unresolved nucleus is significantly lower, in most objects, than that measured at $\sim 0.5$\micron. 
The exceptions are NGC\,4593 and NGC\,3227, both 
of which have relatively low optical polarization and comparably low polarization at 2\micron. The wavelength dependence of $p(\lambda)$ can be attributed, at optical 
wavelengths, to dilution by the reddened, direct AGN (accretion disk) continuum, with a contribution from unpolarized stellar emission (which is required to account for the 
local increases in $p$ over the broad emission lines; \citealt{2004MNRAS.350..140S}). An old stellar population would also dilute the polarization at 2\micron. 
However, it seems likely that the dominant effect in the unresolved nucleus is dilution by hot dust emission. As already 
noted in Section~\ref{scatgeom}, black body components  attributed to hot  dust emission have been identified in several Seyfert galaxies, 
including Mrk\,1239, Mrk\,766 and NGC\,4593. In the latter two cases, in particular, our aperture polarimetry reveals behavior consistent with
the presence of a diluting component associated with the nucleus: the 2\micron\ 
polarization increases significantly with radius, while the nucleus itself is weakly polarized at a level $\lesssim 0.6$\%, the upper limit for instrumental polarization. 

We note in addition that the wavelength dependence of $p(\lambda)$ also depends on the composition of the scattering medium. 
Dust scattering may well dominate in the PSR for which, if in the Rayleigh regime (grain sizes $\lesssim \lambda$), the scattering cross-section $\propto \lambda^{-4}$. 
Unpolarized diluting sources (e.g., starlight; accretion disk and hot dust emission) will therefore have a much greater effect in the NIR than in the optical band.

\section{Conclusions}\label{cons}

We have measured the nuclear polarization at 2.0\micron\ in six ``polar scattered'' S1 galaxies in an effort to confirm the presence of a compact equatorial scattering 
region 
in these objects. This in turn would support the two-component scattering model for Seyfert galaxies. We find evidence for both equatorial and polar scattering in 
Mrk\,1239, which exhibits the expected signatures of equatorial scattering in the unresolved nucleus, and polar scattering off-nucleus on linear scales of $>100$\,pc. 
In this source, therefore, 
the compact ESR is revealed in NIR scattered light and is clearly resolved from the PSR, which extends beyond the torus. A further implication is that the source of the 
{\em scattered} $2\micron$ emission in the unresolved nucleus is the accretion disk, as in the quasars studied by \citet{2008Natur.454..492K}, rather than torus hot dust 
emission.

The remaining objects exhibit a variety of behavior, suggesting that in general the origin of the 2\micron\ polarization is more complex than envisaged in the simple two 
component scattering model. In Fairall~51 and ESO\,323-G077, the nuclear polarization is evidently dominated by polar scattering, even at 2\micron. This cannot be taken 
as evidence of the {\em absence} of an ESR, it simply implies that the Stokes fluxes produced by this region are smaller than those produced by polar scattering. 
This could be the result of higher extinction to the nucleus, a smaller covering factor, or it could be that the scattered 2\micron\ radiation in these objects comes 
predominantly from hot dust in or near the inner part of the torus which `see's the PSR rather than the ESR. In both NGC\,4593 and Mrk\,766, aperture polarimetry shows an 
increase in the 2\micron\ polarization with radial distance from the nucleus (where, indeed, it is consistent with zero). Conceivably, this behavior could be the result of 
cancellation between the orthogonal polarization components of the two scattering regions within the unresolved nucleus. However, in both cases, there is a large difference 
between the 2\micron\ and optical polarization position angles off nucleus, where polar scattering is expected to dominate. This may not be significant for NGC\,4593, in 
which contamination by interstellar polarization is likely to  be important, but is perplexing in the case of Mrk\,766, which is strongly polarized in the optical. The 
nucleus 
of NGC\,3227 is also not significantly polarized at 2\micron. However,  in this object the surrounding PSR is resolved in deeper imaging polarimetry, to be presented 
elsewhere.
\\

\acknowledgments
We thank the anonymous referee for comments and suggestions that improved this paper. 
Support for Proposal number HST-GO-10160 was provided by NASA through a grant from the Space Telescope Science Institute, which is operated by 
the Association of Universities for Research in Astronomy, Incorporated, under NASA contract NAS5-26555.We acknowledge the usage of the 
HyperLeda database (http://leda.univ-lyon1.fr). This research has made use of the NASA/IPAC Extragalactic Database (NED) which is operated by 
the Jet Propulsion Laboratory, California Institute of Technology, under contract with the National Aeronautics and Space Administration. 
This publication makes use of data products from the Two Micron All Sky Survey, which is a joint project of the University of Massachusetts and 
the Infrared Processing and Analysis Center/California Institute of Technology, funded by the National Aeronautics and Space Administration and 
the National Science Foundation.

\bibliographystyle{apj}
\bibliography{batcheldor}

\end{document}